\documentclass[10pt,a4paper,twoside]{aa}

\usepackage{graphicx}
\usepackage{natbib}
\bibpunct{(}{)}{;}{a}{}{,}
\usepackage{latexsym}
\usepackage{amsmath}
\usepackage{amssymb}
\usepackage{amstext}
\usepackage{color}
\usepackage{psfrag}
\usepackage{txfonts}

\def\cm3{cm$^{-3}$}

\def\12{$^{12}$CO}

\begin{document}
\title{The low frequency radio emission and spectrum of the extended SNR 
\object{W44}:\\
new VLA observations at 74 and 324 MHz}

\author {G. Castelletti\inst{1}
\thanks{ Post-Doc Fellow of CONICET, Argentina}, 
G. Dubner\inst{1}
\thanks{ Member of the Carrera del Investigador Cient\'\i fico of CONICET,
Argentina},
C. Brogan\inst{2}, \and
N.~E. Kassim\inst{3}
}

\offprints{G. Castelletti}
\institute {Instituto de Astronom\'{\i}a y  F\'{\i}sica del Espacio (IAFE),
CC 67, Suc. 28, 1428 Buenos Aires, Argentina\\
             \email{gcastell@iafe.uba.ar}
\and
National Radio Astronomy Observatory, 520 Edgemont Road, Charlottesville, VA 22903
\and
Remote Sensing Division, Code 7213, Naval Research Laboratory, 4555 Overlook Avenue, SW, Washington DC } 
   \date{Received <date>; Accepted <date>}

  \abstract
   {} 
  {We present new Very Large Array (VLA) radio images at 74 and 324 MHz of the 
  SNR \object{W44}. The VLA images, obtained with unprecedented angular 
  resolution and sensitivity for such low frequencies  
  have been used in combination with existing 1442 MHz radio data, 
  \it Spitzer \rm IR data, and \it ROSAT \rm and \it Chandra \rm X-ray data to
  investigate morphological and spectral continuum properties of this SNR.
  }
  {The observations were carried out with the VLA simultaneously at 74 and 
  324 MHz in the A and B configurations and at 324 MHz in the C and D 
  configurations. The radio continuum spectral index distribution was 
  derived through direct comparison of the combined data at 74, 324, and 1442 
  MHz. In addition, to isolate and identify 
  different spectral components, tomographic spectral analysis was performed.
  }
  {The spatially resolved spectral index study revealed 
  that the bright filaments, both around and across the 
  SNR, have a straight spectrum between 74 and 1442 MHz, with
  $\alpha\sim -0.5$, with two clear exceptions: a short portion of the SNR 
  limb to the southeast, with $\alpha$ varying between 0 and +0.4
  and a bright arc to the west where the spectrum breaks around 300 MHz and 
  becomes concave down. We conclude that at the shell and along the internal 
  filaments, the electrons responsible for the synchrotron emission 
  were accelerated at the shock according to a simple diffusive shock model. 
  The positive spectrum corresponds to a location where the SN shock is 
  running into a molecular cloud and the line of sight intersects the 
  photo dissociation region of an HII region and a young stellar object is 
  present. Such spectral inversion is a classic signature of
  thermal absorption, either from ionized gas in the postshock region, from 
  the HII region itself, or both. 
  The curved spectrum on the westernmost bright arc is explained 
  as the consequence of strong post-shock densities and enhanced magnetic 
  fields after the interaction of the SN shock with a coincident molecular 
  cloud. 
  The comparison of the 324 MHz image with a 4.5 $\mu$m IR image obtained 
  with \it Spitzer \rm underscored an impressive correspondence between  
  emission both to the north and west sides of \object{W44}, while 
  the comparison with \it ROSAT \rm and \it Chandra \rm images 
  confirm that the synchrotron radio emission surrounds the thermal X-ray 
  radiation. 
  }
  {}

\titlerunning{Low-frequency observations of SNR \object{W44}}
\authorrunning{\textsc{Castelletti et al.}}

\maketitle

\section{Introduction}

The interaction of supernova remnant (SNR) shocks with inhomogeneous
interstellar gas strongly affects the spatial and temporal evolution of
SNRs.  The density distribution of the ambient circumstellar material
(CSM) and interstellar medium (ISM) can influence the formation of a
variety of SNR morphologies, including specific sub-classes of SNRs
such as ``thermal-X-ray composites'' \citep[also called
mixed-morphology SNRs,][]{rho98}, or remnants with bilateral symmetry
\citep{gae98}.  In this respect, sensitive high-resolution radio
observations of SNRs are particularly useful for studying the physics
of shock waves traveling through nonuniform environments in the ISM.

On the other hand, continuum radio spectra made accurate by extension
to long wavelengths provide sensitive and often unique means of
distinguishing between different physical processes taking place
either within SNRs (e.g. shock acceleration), in their immediate
surrounding, or in the ISM along the line of sight \citep[e.g. thermal
absorption,][]{lac01,bro05}. Based on accurate spectral continuum index 
studies it is
possible to constrain shock acceleration theories that predict subtle
variations in both spatially resolved \citep{and93} and integrated
spectra \citep{rey92}. In addition, the information provided by low
frequency observations of SNRs is a powerful complement to higher
frequency radio, X-ray, and infrared (IR) data to study the complex
processes associated with SNR shocks \citep{bro05}.

\object{W44} (\object{G34.7$-$0.4}) is a bright radio SNR 
($S{_\mathrm{1GHz}}$$\sim$230 Jy) with an
asymmetric morphology, about 0\fdg5 in size, and located in a complex
region of the inner Galactic plane rich in both thermal and nonthermal
sources. \object{W44}  is an archetype of the ``mixed-morphology'' class of
remnants, characterized by a highly filamentary radio shell and
centrally concentrated thermal X-ray emission.  Morphological studies
of this remnant in the optical band reveal Balmer-dominated, [NII],
and [SII] filaments from radiative shocks, with some concordance with
the radio emission \citep{rho94, gia97}.

This SNR is also of interest because it constitutes one of the few
demonstrated cases of a SNR-molecular cloud interacting system.
\citet{set98} observed six giant molecular clouds that appear to be
partially surrounding the remnant. Higher spatial resolution $^{12}$CO
observations of the full extent of \object{W44} presented by
\citet{set04} revealed that some of these 
clouds (at  v$_{LSR}$$\sim$48 km~s$^{-1}$) are physically 
interacting with the remnant on its southeastern and western sides.
These observations also show that the shocked molecular gas
is confined to compact ($\sim$1.5 pc) cores located adjacent to bright
radio filaments or knots.  The physical interaction of the blast wave
with the clumpy ISM is also supported by the detection
of bright OH (1720 MHz) masers  
\citep[at LSR velocities between $\sim$43 and $\sim$47 km~s$^{-1}$,][ and 
references therein]{hof05}, and IR cooling lines from H$_{2}$ \citep{rea06}.

The associated pulsar \object{PSR~B1853+01} is located inside the
\object{W44} shell, about 9$^{\prime}$ south of the geometric center
of the remnant. A characteristic age of about 2 $\times$ 10$^{4}$
years has been estimated for this pulsar \citep{wol91}.  \citet{fra96}
and \citet{pet02} report the detection of a radio and X-ray nebula
powered by an associated pulsar wind.

It has also been proposed that the EGRET source 3EG~1856+0114 is
associated with \object{W44} \citep{es96}.  The 95\% confidence
contour for this $\gamma$-ray source coincides with the southeastern
sector of the remnant \citep{tho96}.  \citet{dej97} explained the
observed GeV $\gamma$-rays in terms of relativistic bremsstrahlung and
inverse Compton scattering produced by a power-law distribution of
relativistic electrons injected by the pulsar \object{PSR~B1853+01}.
While this constitutes a possible explanation, a significant part of
the $\gamma$-ray flux observed may also be due to SNR shock-cloud
interactions \citep{fat05}. No $\gamma$-ray radiation at TeV energies
was detected towards \object{W44} using either the Whipple \citep{les95}
or CANGAROO \citep{buc98} telescopes, suggesting a spectral cutoff
between GeV and TeV energies \citep{row00}.

A distance of about 3 kpc was first suggested for \object{W44} by \citet{rad72}
and \citet{cas75} based on H$\,$\textsc{i} absorption measurements. From the 
afore mentioned molecular studies revealing associated gas with LSR 
velocities between approximately 43 and 48 km~s$^{-1}$
and using a flat rotation curve with the solar constants of
R$_{0}$=7.6 kpc \citep{eis05} and $\Omega$$_{0}$=27.2 km~s$^{-1}$~kpc$^{-1}$ 
\citep{fea97}, a kinematical distance of 2.9 $\pm$ 0.2 kpc can be derived 
for W44. Thus \object{W44} is probably an object of the Sagittarius arm
consistent with it being a core-collapse SNe from a massive progenitor star. 

In this paper we present the highest resolution and sensitivity images
of the SNR \object{W44} yet obtained at low radio frequencies. These
images were generated from multiple-configuration Very Large Array
(VLA)\footnote{The Very Large Array of the National Radio Astronomy
Observatory is a facility of the National Science Foundation operated
under cooperative agreement by Associated Universities, Inc.}
observations at 74 and 324 MHz carried out in 2002 and 2003. Based on
the comparison of the new images with existing VLA data at 1442 MHz,
we have performed the first detailed study of the radio continuum spectral
properties of \object{W44}, searching for changes in spectral
index as a function of frequency and position. In addition, special attention 
was paid to the comparison between our high resolution radio images and
the best existing infrared data from {\em Spitzer} and in
X-rays from {\em ROSAT} and {\em Chandra}.

\section{Observations and Data reduction}

We have performed low frequency observations of the SNR \object{W44}
at 74 and 324 MHz using multiple-configurations of the VLA.
Table~\ref{observations} summarizes various observing details.  Data
at both frequencies were obtained in spectral-line mode to enable
radio-frequency interference (RFI) excision and to mitigate bandwidth
smearing.  The observations at 74 MHz and 324 MHz were performed with
64 and 32 channels, respectively.  A model image of \object{Cygnus A}
was used to correct the bandpass\footnote{See
http://lwa.nrl.navy.mil/tutorial/ to obtain source models in FITS
format.}. The absolute amplitude scale were set using \object{Cygnus
A} at 74 MHz and \object{3C~48} at 324 MHz.  Also, for the 324 MHz
data, regular observations of the secondary calibrator
\object{1822$-$096} were used for initial phase and amplitude
calibration.  At both frequencies, the data from each array were fully
reduced and imaged separately using the NRAO Astronomical Image
Processing Software (AIPS) package.  After initial calibration the
data at 74 and 324 MHz were averaged down to fourteen and eight
channel spectral resolution data cubes, respectively, before imaging.

Two additional difficulties associated with low frequency radio
observations are the wide-field imaging requirement imposed by the
large field of view (the primary beam of the 324 and 74 MHz data are
2\fdg5 and 11\fdg5, respectively), and ionospheric-based phase
variations that decorrelate phases on longer baselines, especially at
74 MHz \citep{kas93}.  We therefore employed wide-field imaging as
implemented in the AIPS task IMAGR, based on a
pseudo-three-dimensional multifacet algorithm to deal with the
non-coplanarity of the visibilities \citep{cor92}.  Secondly, to
remove ionospheric-induced phase distortions we employed several loops
of self-calibration to the data from each configuration separately at
74 and 324 MHz \citep{cor99}. We note that angle-invariant
self-calibration (as currently implimented in AIPS) is not sufficient
to correctly compensating for ionospheric effects across the full
field of view, especially at 74 MHz \citep{cot04}. However,
self-calibration is adequate in the special case where the target
source is located at the phase center and is the brightest source in
the field, as is this case with \object{W44}.

The final calibrated 324 MHz visibility data from the A, B, C, and D
configurations were combined into a single {\em uv} data set and
imaged using a multi-scale CLEAN algorithm with four different scale
sizes. For the final image, all facets were reassembled into one large
field with the task FLATN within AIPS, resulting in a single image
with an angular resolution of $13\farcs 3\times 13\farcs 2$,
PA=$-4\fdg7$, and an rms noise of 5 mJy~beam$^{-1}$.

Data reduction and imaging at 74 MHz were performed following the
procedures outlined above\footnote{see at
http://lwa.nrl.navy.mil/tutorial/ for a full
description on 74 MHz VLA data reduction.}. For the A and B
configuration 74 MHz data, an initial phase-only self-calibration was
conducted using the B-configuration 324 MHz image as the starting model. The A
and B configuration 74 MHz data sets were then phase-only self-calibrated;
the B configuration data also required a final amplitude self-calibration. 
The individually calibrated A and B data sets were then
concatenated, and followed by several more rounds of self-calibration.
The final image has a synthesized beam of $39\farcs 1\times 35\farcs
9$, PA=$-47^{\circ}$ and an rms noise of 65 mJy~beam$^{-1}$.

A concern with self-calibrated images is that they can lose
absolute position information subject to the arbitrary position of the 
starting model, which in this case may also be subject to ionospheric
refraction.  In order to check the astrometry in our 74 and 324 MHz
images, we compared the position of several small-diameter sources
detected in these images with their positions from the NRAO VLA Sky
Survey (NVSS). This catalog has an astrometric accuracy better than
1$^{\prime\prime}$ in both RA and Dec for bright sources
\citep{con98}. We found and corrected for an average ionospheric induced 
refraction of
0.63$^{\mathrm{s}}$$\pm$ 0.12$^{\mathrm{s}}$ in RA and $9\farcs1\pm 1\farcs6$ 
in Dec at 74 MHz and 0.14$^{\mathrm{s}}$$\pm$ 0.05$^{\mathrm{s}}$ in RA and 
$3\farcs52\pm 0\farcs93$ in Dec at 324 MHz.

A high resolution VLA image of \object{W44} at 1442 MHz was produced
from interferometric data taken in 1984 and 1985 \citep{jon93} plus short
spacing information from the Bonn 100 m 1408 MHz Survey \citep{rei90}.  
This 1442 MHz image recovers a total flux
density of S$_{1.4\,\mathrm{GHz}}$=300 $\pm$ 7 Jy and has a 
resolution and rms noise of 
23$^{\prime\prime}$ $\times$ 19$^{\prime\prime}$ and 4 mJy~beam$^{-1}$, 
respectively.

To study the spectral properties of \object{W44} (\S4), special care
was taken to match the {\em uv} coverage for the 74, 324, and 1442 MHz
VLA data. This was achieved by applying specific weighting schemes to
the 324 MHz data in order to provide the same resolution as the 74 and
1442 MHz images, respectively. Similarly, we re-imaged the 1442 MHz
data so that they matched the 74 MHz observations. Primary beam
corrections were applied to the new 74 and 324 MHz images before
estimating flux densities.

\begin{table}
\renewcommand{\arraystretch}{1.0}
\centering
\vspace{0.26cm}
\caption{Observational summary}
\begin{tabular}{lccc}  \hline \hline
Observing      & VLA &   Bandwidth & Integration \\
dates          & Config. &   (MHz) &       time (hr)\\
\hline
\multicolumn{4}{c}{74 MHz Parameters} \\
\hline
2003 Aug 31  & A   & 1.562 & 7.5  \\
2003 Sep 1  & A   & 1.562 & 2.6   \\
2002 Jun 15 & B   & 1.562 & 6.0   \\
\hline
\multicolumn{4}{c}{324 MHz Parameters} \\
\hline
2003 Aug 31  & A   & 3.125 & 7.5  \\
2003 Sep 1  & A   & 3.125 & 2.6   \\
2002 Jun 15 & B   & 3.125 & 6.0   \\
2002 Dec 14 & C   & 3.125 & 4.0   \\
2003 Mar 16  & D   &3.125 & 6.0   \\ \hline
\label{observations}
\end{tabular}
\end{table}

\section{Results}
\subsection {VLA low frequency images of \object{W44}}

\begin{figure*}[h]
\centering
\includegraphics[]{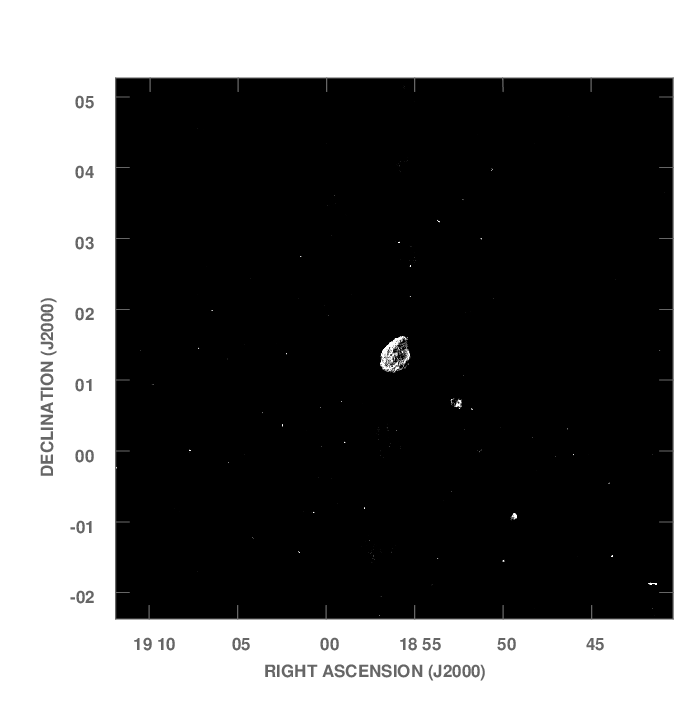}
\caption{
This image shows a portion of the primary beam of the VLA at 74 MHz
centered on the SNR \object{W44}, after the combination of A- and
B-configuration data. The
image displayed does not include primary beam correction. 
The field covers $\sim$7.5 square degrees at an angular
resolution of 39$^{\prime\prime}$ $\times$ 36$^{\prime\prime}$ 
(P.A.=$-$47$^{\circ}$). The rms noise is 65
mJy~beam$^{-1}$.
\label{74}}
\end{figure*}

\begin{figure}[ht]
\centering
\includegraphics[]{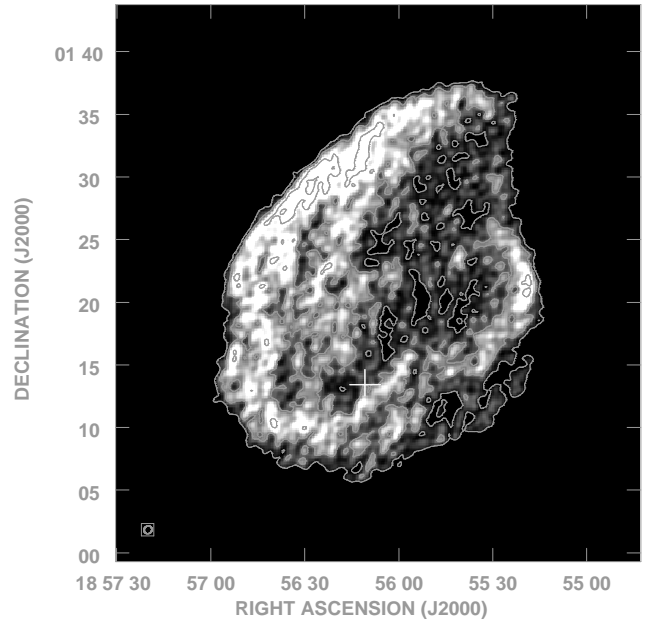}
\caption{
Radio continuum emission at 74 MHz from the SNR \object{W44}.  The
grayscale flux is linear ranging from 260 to 700 mJy~beam$^{-1}$. The
rms noise level is 65 mJy~beam$^{-1}$, and the contours are traced at
multiples of the 4$\sigma$ noise level, 0.260 $\times$ (1, 1.76, 3.30,
4.46) Jy~beam$^{-1}$.  The synthesized beam shown at the bottom left
corner is 39$^{\prime\prime}$ $\times$ 36$^{\prime\prime}$.  The 
white plus sign to the south of \object{W44} indicates the position of the
\object{PSR~B1853+01}.
\label{74only}}
\end{figure}

Figure~\ref{74} shows our $\sim 37\arcsec$ resolution VLA 74 MHz image
of a portion ($\sim\,$7.5 square degrees) of the large field of view
centered on the SNR \object{W44}. Note that while sources are detected
across the full field-of-view, only those located near the field
center are correctly imaged by the angle-invariant self-calibration
technique implimented here. A close-up view of the radio emission at
74 MHz from \object{W44} is shown in Fig.~\ref{74only}. This is the
first sub-arcminute resolution image of this source ever obtained
below 100 MHz. Though it is not sensitive to structures larger than
$\sim 37\arcmin$, it is sufficient to recover all information 
for this source about 30$^{\prime}$ in size. 
Moreover, the fact that no negative bowl is apparent around the SNR at
this frequency confirms that most of the large-scale structure is recovered.
However, since \object{W44} is almost at the limit of the largest imaged 
structures, some flux might still be missing from the 74 MHz image. 

\begin{figure*}[ht]
\centering
\includegraphics[]{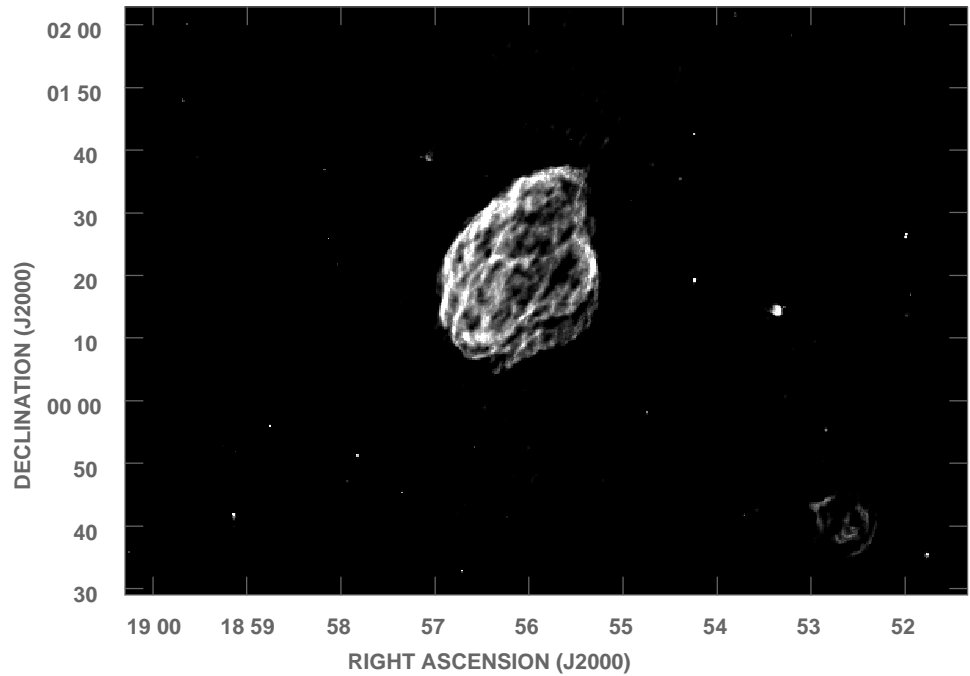}
\caption{
An image of the field of view around the SNR \object{W44} at 324 MHz.
The image displayed does not include primary beam correction.
The grayscale varies linearly between 15 and 70 mJy~beam$^{-1}$. The
angular resolution is about 13$^{\prime\prime}$ $\times$
13$^{\prime\prime}$ with a position angle of $-4\fdg7$ and the
sensitivity level is 5.0 mJy~beam$^{-1}$. The source located near the
bottom right corner ($\sim$ 18 52 30, +00 20) is the SNR \object{Kes~79}.
\label{324}}
\end{figure*}

\begin{figure*}[h]
\centering
\includegraphics[]{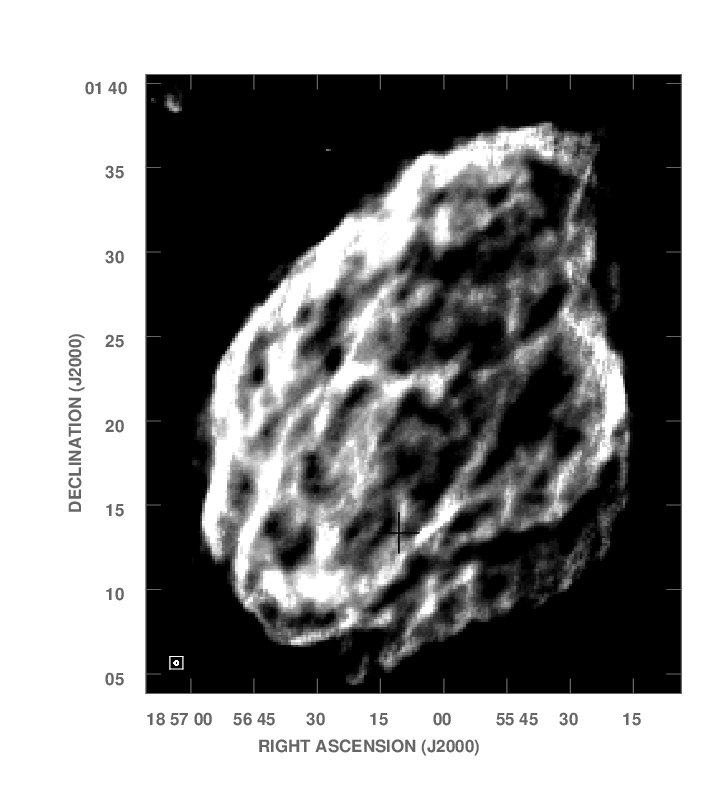}
\caption{
Image of SNR \object{W44} at 324 MHz. The grayscale is linear ranging 
from 20 to 60 mJy~beam$^{-1}$. The rms noise
level is 5.0 mJy~beam$^{-1}$. The synthesized beam shown at the bottom left 
corner is 13$^{\prime\prime}$ $\times$ 13$^{\prime\prime}$.
The black plus sign indicates the position of the \object{PSR~B1853+01}.
\label{324only}}
\end{figure*}

In Fig.~\ref{324} we present the new VLA image of \object{W44} at 324
MHz with a resolution of $13\arcsec$ over a field of view of $\sim
2.5$ square degrees.  A close-up view showing the detailed radio
morphology of \object{W44} at 324 MHz is shown in Fig.~\ref{324only}.
The present image has a beam size 300 times better 
and is 80 times more sensitive than any previously published 330
MHz image of \object{W44} \citep{kas92}.  The combination of all four
VLA array configurations is sensitive to structures between 13$\arcsec$ and
1$^{\circ}$ in size, ensuring complete sampling of all the spatial
scales of the \object{W44} emission. The high dynamic range achieved
in these new observations has served to reveal new details previously
unnoticed in \object{W44}, especially extreme filamentary and clumpy
emission that is immersed in more tenuous diffuse emission.

At low radio frequencies the remnant has the appearance of a
noncircular shell elongated in the southeast-northwest direction, with
an average diameter of $\sim$30$^{\prime}$ (or 25  pc at the distance of
2.9 kpc adopted in this paper). The brightest emission occurs along the eastern 
boundary, a region where previous observations have indicated the presence of
molecular clouds probably interacting with \object{W44}.
In the 324 MHz image (Fig.~\ref{324only}) it is apparent
that the east limb is not sharp. On the contrary, faint emission is
clearly detected at a 5$\sigma$ level extending about 1$^{\prime}$
beyond the bright rim. A similar weak radio halo was observed along
the eastern side of the SNR \object{Puppis~A} \citep{cas06}, a
direction where \object{Puppis~A} is encountering a molecular
cloud as is the case for \object{W44}.  
To the west, the most outstanding feature is the short bright
arc located near $18^{\mathrm{h}}\,55^{\mathrm{m}}\,20^{\mathrm{s}}$,
$01^{\circ}\,22^{\prime}$ (J2000).  The rest of the boundary is
more diffuse in appearance.

Based on the new radio observations we have estimated integrated flux
densities of 634 $\pm$ 70 Jy at 74 MHz and 411 $\pm$ 50 Jy at 324 MHz
for \object{W44}, where the attenuation of the primary beam has been
corrected for these estimates. The quoted errors take into account the
statistical errors as well as uncertainties in the true extension of
the SNR.   

\begin{figure}[h]
\centering
\includegraphics[]{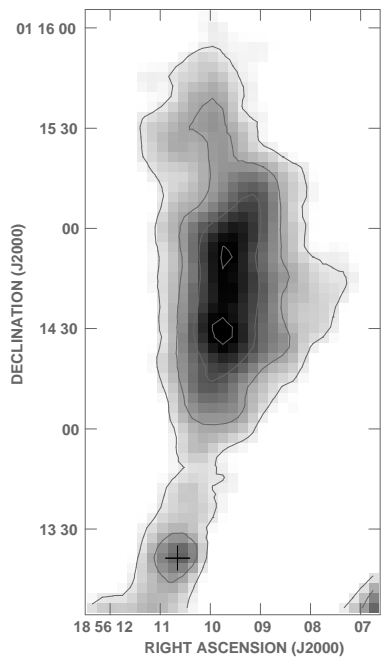}
\caption{A close-up image of the pulsar wind nebula around
PSR~B1853+01 at 324 MHz.  The position of the pulsar is indicated by
the plus sign. The grayscale runs from 34 to 56 mJy~beam$^{-1}$ and the
contour levels are 0.036, 0.042, 0.049, 0.056 mJy~beam$^{-1}$.
\label{pwn}}
\end{figure}

In the new sensitive 324 MHz image, the low frequency counterpart of
the pulsar wind nebula (PWN) around \object{PSR~B1853+01} discovered
at higher radio frequencies and in X-rays \citep{fra96,pet02}, is
clearly detected (small white cometary feature extending to the north of the
pulsar position indicated by the plus sign in Fig.~\ref{324only}). An
enlargement of the emission at 324 MHz associated with this PWN is
shown in Fig.~\ref{pwn}.  Similar to the 1.4 GHz image shown by
\citet{fra96}, at 324 MHz the pulsar is located at the apex of an
elongated structure $\sim 2\farcm5$ in extent. 
The two maxima observed inside the PWN, where the intensity peaks up to 56
mJy~beam$^{-1}$, were previously reported by \citet{jon93} as discrete sources.
The integrated flux density of the PWN at 324 MHz is
$0.35\pm 0.12$ Jy.  Because of the relatively lower sensitivity and
angular resolution, combined with the typically flat radio spectra 
of all PWN, the W44 PWN is not apparent in the 74 MHz image.

\section{The radio spectral index behavior}
\subsection{Integrated radio spectrum}
We have used the new estimated total flux densities at 74 and 324 MHz
together with data from the literature to determine the mean spectral index of
\object{W44}.  To construct accurate spectra of individual SNRs it is
necessary that all flux densities be tied to the same
absolute flux density scale.  In Table~\ref{globalW44} we list all
integrated flux density estimates for \object{W44} between 22 and
10700 MHz.  Most of the measurements over the frequency range between
408 and 10700 MHz were extrapolated to the Baars et al. (1977) absolute flux
scale for consistency.  In some cases an estimate of the correction
factor was not available because the original reference did not list
the assumed flux densities of the primary calibrators.  Below 408 MHz
($\log\,\nu \sim$ 2.6), where the systematic error of the Baars et al.
scale is more than $\sim$ 5\%, no scaling was applied.
These last values have been nevertheless used for the spectral
fitting since they do not show a large spread.

\begin{figure}[]
\centering
\includegraphics[]{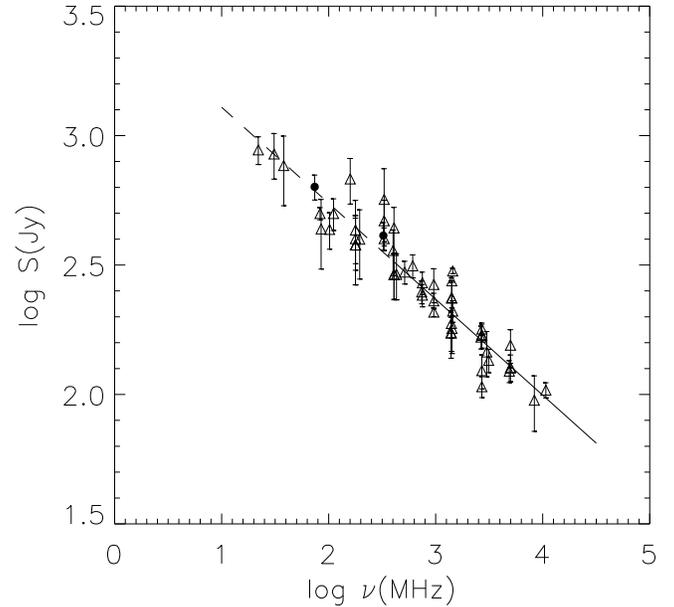}
\caption{Integrated radio continuum spectrum of \object{W44} obtained
from the flux density values listed in Table~\ref{globalW44}.  The
filled symbols correspond to data from the new VLA measurements at 74
and 324 MHz, while the rest of the values were taken from the literature
and, where possible, brought onto a single flux density scale. 
The linear fit to all of the flux density values yields a 
spectral index $\alpha$=$-$0.37 $\pm$ 0.02.
(S$_{\mathrm{\nu}}\,\propto\,\nu^{\alpha}$) and is shown by the
line. The dashed part of the line indicates the range of frequencies over 
which no
correction to bring the measured values to the \citet{baa77} scale was
available.
\label{spectrum}}
\end{figure}

In Fig.~\ref{spectrum} we show a plot of the spectrum for \object{W44}
in which our new flux density determinations at 74 and 324 MHz are
indicated by filled circles.  A single power law slope
$\alpha$=$-$0.37 $\pm$ 0.02 (S$_{\mathrm{\nu}}\propto\nu^{\alpha}$)
adequately fits the flux densities measured over four decades in
frequency without any obvious break. This integrated spectral index is
quite flat for a typical shell type SNR, and agrees well with earlier
estimates presented by \citet{kas89a, kas89b, kas92} who derived a
spectral index between $-$0.4$\,<$ $\alpha$ $<$$\,-$0.3.  This result
confirms \citet{kas89a}'s contention that the integrated spectrum of
\object{W44} remains a power law to well below 100 MHz. This behavior
is in contrast to a number of other SNRs that exhibit a spectral turnover
due to free-free absorption \citep[e.g. W49B and
3C391,][]{lac01,bro05}. The fact that the new 74 MHz integrated flux falls
nicely on the power law confirms that our 74 MHz image is missing little, if
any emission on the largest spatial scales.

\begin{table*}[h!]
\renewcommand{\arraystretch}{1.0}
\centering
\caption{Integrated flux densities on the SNR \object{W44}}
\label{globalW44}
\vspace{0.26cm}
{\begin{tabular}[width{22cm}]
{llllll}   \hline\hline
Frequency  & Scaled flux  &  References &Frequency  & Scaled flux  &  References\\
 (MHz)  &  density (Jy)    &   & (MHz)  &  density (Jy)\\ \hline
 22\dotfill & 881$\pm$108$^{\mathrm{(a)}}$ & \citet{rog86}& 750\dotfill & 270 $\pm$ 27 & 
 \citet{pau66} \\
 30.9\dotfill& 849 $\pm$ 170$^{\mathrm{(a)}}$  &\citet{kas89b} & 960\dotfill & 208 $\pm$  8
 &
 \citet{har62} \\
 38 \dotfill & 766 $\pm$ 230$^{\mathrm{(a)}}$  &\citet{kas89b} &
 960\dotfill & 230 $\pm$  16$^{\mathrm{(c)}}$  &
 \citet{kuz62} \\
 74\dotfill &  634 $\pm$ 70$^{\mathrm{(b)}}$ & This work & 960\dotfill & 266 $\pm$ 40 &
 \citet{wil63} \\
 83\dotfill &  500 $\pm$ 27$^{\mathrm{(a)}}$ & \citet{kov94} &1390\dotfill & 173  $\pm$ 35 &

 \citet{wes58} \\
 85\dotfill &  436 $\pm$ 131$^{\mathrm{(a)}}$ &\citet{mil58} &1400\dotfill & 188  $\pm$ 23 &

 \citet{pau66}\\
 102\dotfill & 434 $\pm$ 70$^{\mathrm{(a)}}$ & \citet{kov94} &1400\dotfill & 173 $\pm$ 26 &
 \citet{kel69}  \\
 111\dotfill & 500 $\pm$ 70$^{\mathrm{(a)}}$ & \citet{kov94}  &1410\dotfill & 236  $\pm$ 47
 &
 \citet{sch63} \\
 159\dotfill & 680 $\pm$ 122$^{\mathrm{(a)}}$ & \citet{edg58} &1410\dotfill & 236  $\pm$ 35
 &
 \citet{bea69} \\
 178\dotfill & 432 $\pm$ 130$^{\mathrm{(a)}}$ &\citet{ben63} & 1414\dotfill & 274.7 $\pm$ 0.
 4 &
 \citet{alt70}\\
 178\dotfill & 400 $\pm$ 80$^{\mathrm{(a)}}$ & \citet{ben63} & 1420\dotfill & 180  $\pm$ 36$
 ^{\mathrm{(c)}}$ & \citet{les60} \\
 178\dotfill & 378 $\pm$ 113$^{\mathrm{(a)}}$   &\citet{kel69} & 1442\dotfill & 
 210 $\pm$ 20 & \citet{gia97}\\
 178\dotfill & 378 $\pm$ 113$^{\mathrm{(a)}}$  &\citet{hol69} & 1442\dotfill &300
 $\pm$ 7 & This work \\
 195\dotfill & 398 $\pm$ 119$^{\mathrm{(a)}}$   &\citet{kun67} & 2650\dotfill & 167
 $\pm$ 17 & \citet{bea69}\\
 324\dotfill & 411 $\pm$ 50$^{\mathrm{(b)}}$ &  This work &  2695\dotfill & 170.3 $\pm$ 0.4 & \citet{alt70} \\
 330\dotfill & 469 $\pm$ 94$^{\mathrm{(a)}}$ & \citet{kas92} & 2700\dotfill & 107 $\pm$ 10&
 \citet{alt70} \\
 330\dotfill & 400 $\pm$ 40$^{\mathrm{(a)}}$ & \citet{gia97} &2700\dotfill & 169  $\pm$ 18&
 \citet{wil73}\\
 400\dotfill & 360 $\pm$ 72$^{\mathrm{(a)}}$ & \citet{dav65} &2700\dotfill & 179  $\pm$ 10 &
  \citet{vel74}\\
 408\dotfill & 291 $\pm$ 58 & \citet{lar61}  &2700\dotfill & 124 $\pm$ 19 &
  \citet{mil74}\\
 408\dotfill & 440 $\pm$ 88 & \citet{kes68}  &3000\dotfill & 146  $\pm$ 29 & \citet{sch63} \\
 408\dotfill & 290 $\pm$ 58 & \citet{clar75}  & 3125\dotfill & 135 $\pm$ 14 &\citet{kuz60}\\
 430\dotfill & 291 $\pm$ 59 & \citet{kun67} & 4875\dotfill & 123 $\pm$ 12 &\citet{dow80} \\
 430\dotfill & 567 $\pm$ 178 & \citet{dic75}  & 5000\dotfill & 127  $\pm$ 15 & \citet{kun69}\\
 513\dotfill & 297 $\pm$ 30$^{\mathrm{(c)}}$ & \citet{kuz62} & 5000\dotfill & 155  $\pm$ 23 & \citet{mil69}\\
 610\dotfill & 314 $\pm$ 32 & \citet{mor65} & 5000\dotfill & 126.7 $\pm$ 0.4 &
  \citet{alt70} \\
 740\dotfill & 249 $\pm$ 15$^{\mathrm{(c)}}$ & \citet{kuz62} & 8350\dotfill &  95  $\pm$ 23$^{\mathrm{(c)}}$&
\citet{hol66}\\
 750\dotfill & 242 $\pm$ 24 & \citet{kel69}  & 10700\dotfill& 104  $\pm$  7 &
\citet{kun72}\\
\hline
\end{tabular}}
\begin{list}{}{}
\item[$^{\mathrm{(a)}}$] No correction to \citet{baa77} scale was applied.
\item[$^{\mathrm{(b)}}$] Flux density scale from VLA Calibrator Manual,
http:/www.aoc.nrao.edu/$\sim$gtaylor/calib.html.
\item[$^{\mathrm{(c)}}$] The correction factor was not available.
\end{list}
\end{table*}

\subsection{Local variations in the radio spectral index}

We aim to discern whether the spectrum is a global
property of this SNR, or if it is locally affected by the shock-ISM
interaction and/or the presence of the pulsar in its
interior. To investigate this, we performed a detailed study of the
variations in the spectrum as a function of frequency and position
within the remnant. For this analysis we have used the new high
resolution images at 74 and 324 MHz together with the reprocessed VLA
data at 1442 MHz.

To accurately determine the spectral index distribution based on
interferometric radio images, identical beam size and shape at the
different pairs of frequencies is required. As mentioned in \S2,
we applied appropriate {\em uv}-tapering, in order to match the range
of spatial scales measured at each frequency.  In addition, to avoid
any positional offsets, the images were aligned and interpolated to
identical projections (field center, pixel separation, etc).  The
final common resolution is 40$^{\prime\prime}$ $\times$ 36$^{\prime\prime}$,
PA=$-$62$^{\circ}$ for the 74/324 MHz comparison and
14$^{\prime\prime} \times 14^{\prime\prime}$,
PA=$-$45$^{\circ}$ for the 324/1442 MHz pair.

The spectral study was carried out by direct comparison of
the matched images at the three frequencies and 
by constructing tomographic images. The
different procedures ensure that the observed spectral features are
robust, and unaffected by any artifacts or biases in the databases
used for the comparison, like zero level differences, smooth
background variation, etc.

In spite of the relatively good fidelity attained in the new images, some 
residual 
small-scale fluctuations, probably not real, were apparent at the highest
spatial resolution traced in the spectral index maps. We therefore preferred to further 
smooth all three frequency maps to a 50$\arcsec$ resolution, thus 
avoiding any small scale variation that could mask the main spectral features. 
We based our analysis on spectral comparisons derived from these smoothed 
databases.

\subsubsection{Spectral index maps}

\begin{figure*}[]
\centering
\includegraphics[]{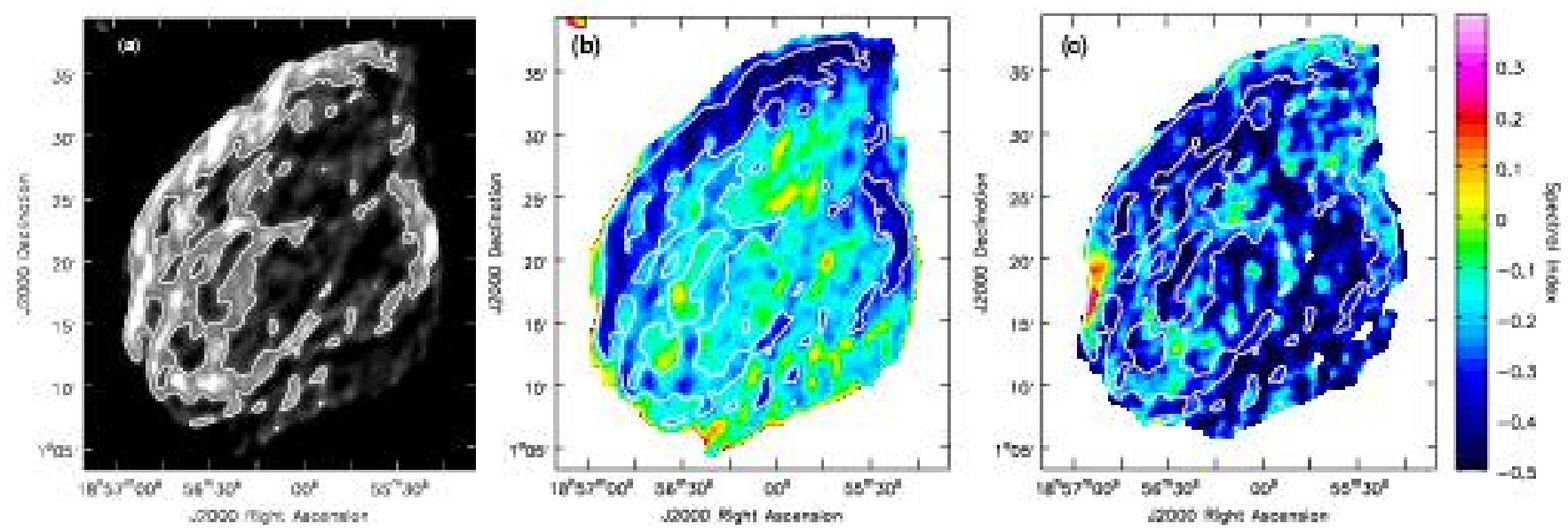}
\caption{\bf (a) \rm 324 MHz image of \object{W44} with 25$\arcsec$
resolution showing the location of the brightest filaments. \bf (b) \rm
Spectral index map between 324 and 1442 MHz (50$\arcsec$ resolution).
\bf (c) \rm Spectral index map between 74 and 324 MHz (50$\arcsec$
resolution). The 0.15 Jy~beam$^{-1}$
contour from the 25$\arcsec$ resolution 324 MHz image is included on each
panel to facilitate the comparison between spectral continuum and total power
features. Only regions with flux densities greater than 4$\sigma$ were
used to create the spectral index maps in (b) and (c). Both spectral
index maps have the same color scale (displayed to the right).
\label{brogan}}
\end{figure*}

Figure~\ref{brogan}a  shows a 324 MHz continuum image of the SNR
\object{W44} smoothed to 25$\arcsec$ resolution for comparison, while 
Fig.~\ref{brogan}b and Fig.~\ref{brogan}c
display the spatial spectral index distribution in the remnant
obtained from the direct ratio of the matched 324/1442 MHz and
74/324 MHz data, at the angular resolution of 50$\arcsec$. To construct these
spectral index maps the images were clipped at the 4$\sigma$ level of their 
respective noise levels. Since, as noted in \S4.1, the global spectrum of
\object{W44} remains a power law to well below 100 MHz, we use the
same scale in Fig.~\ref{brogan} for both frequency pairs in order to
easily identify continuum spectral changes that depend on both position and
frequency (spectral curvature).

In Fig.~\ref{brogan}b it is readily apparent that the brightest radio 
emission all along the periphery of \object{W44} has a spectral index
$\alpha_{324}^{1442}$$\sim\,$$-$0.4, $-$0.5, compatible with a first-order 
Fermi mechanism operating at the shock. In addition, the bright filaments 
observed across the SNR have a similar $\alpha$, in contrast with the slightly 
flatter diffuse interior.

In the spectral index map traced between 74 and 324 MHz (Fig.~\ref{brogan}c)
the most striking feature is the short band seen on the southeastern limb (near
$18^{\mathrm{h}}\,56^{\mathrm{m}}\,50^{\mathrm{s}}$,
$01^{\circ}\,17^{\prime}$), where the spectral index becomes
positive, with $\alpha$ varying between 0 and +0.4. This spectral inversion 
occurs at a site where previous studies have demonstrated that the SNR is 
physically interacting with a dense molecular cloud \citep{set04,rea05}. Also, 
as shown below (\S5.1) from 24 $\mu$m and 8 $\mu$m {\it Spitzer} 
observations, this is the location of an HII region 
\citep[named G034.7-00.6 in the HII regions catalog by][centered at 
18$^{\rm h}$ 56$^{\rm m}$ 47.9$^{\rm s}$, 01$^{\circ}$ 17$\arcmin$ 54$\arcsec$ 
J2000]{pal03} 
with the IRAS point source 18544+0112 (identified as a young stellar object)
located on its border. Thus, the spectral inversion can be interpreted as
originating from low frequency free-free absorption at the SNR/molecular cloud
shock boundary. Absorbing ionized gas could have been created by the passage 
of a fast, ionizing J-type shock that passed through the molecular cloud and 
dissociated and ionized 
the gas. A similar spectral inversion was observed by \citet{bro05} in the SNR
\object{3C~391}, where a spectral turnover for frequencies below 100 MHz was 
found at the positions where the SN shock was running into dense molecular gas.
An alternative explanation is that thermal absorption is occuring inside the
boundaries of the coincident HII region, along its periphery where the
thermal electron density might be highest. We note that this localized 
absorption has a negligible influence on the total integrated flux and thus 
has no measurable impact on the integrated continuum spectrum.

From the comparison between Fig.~\ref{brogan}b (324/1442 MHz) and 
Fig.~\ref{brogan}c 
(74/324 MHz) one could have the misleading impression that the spectral 
behavior is reversed from one frequency range to the other.
In fact the spectral index of the limb and bright filaments is the same (near
$\sim\,$$-$0.4 to $-$0.5) in both frequency pairs, and it is the diffuse 
interior that changes to a steeper value in the 74/324 MHz comparison.
This effect, however, needs to be confirmed on the basis of more sensitive 74 
MHz data before advancing an interpretation.

What is readily apparent from the comparison of Fig.~\ref{brogan}b and
Fig.~\ref{brogan}c is that the westernmost arc changes its spectrum from
$\alpha$$_{324}^{1442}$$\sim\,$$-$0.4 to a flatter value
$\alpha$$_{74}^{324}$$\sim\,$$-$0.2.
This feature is also located at a site where the SN
shock is probably running into a dense molecular cloud, where the 
intense near infrared emission (see \S5.1) and  the bright optical filaments
observed in the [SII] and [NII] lines \citep{mav03} suggest the presence of 
radiative shocks. In this case, it is possible that the observed change in the 
radio continuum spectrum is due to absorption by a layer of thermal electrons 
associated with the sheets responsible for the filaments.  

\subsubsection{Tomographic images}
To investigate the fidelity of our radio continuum spectral index maps, we 
also produced a tomographic gallery \citep[see for example][]{kat97}. 
A tomographic map between two images at different frequencies is made by using 
a test spectral index $\alpha_{\mathrm{t}}$ to scale the brightness in
the higher frequency image ($S_{\nu_{1}}$), and then subtracting this
scaled image from the lower frequency image ($S_{\nu_{2}}$). Thus, a
tomographic gallery is obtained by calculating
$S_{t}=S_{\nu_{2}}-(\nu_{2}/\nu_{1})^{\alpha_{\mathrm{t}}}\,S_{\nu_{1}}$,
stepping through a range of test spectral indices
$\alpha_{\mathrm{t}}$. Features which have a spectral index identical
to the test value will disappear in the S$_{t}$ image.  Spatial
components that have different spectral indices will appear as
positive or negative residuals depending upon whether the spectrum is
steeper or flatter than the assumed $\alpha_{\mathrm{t}}$ value. Such
a procedure allows us to trace fine-scale spectral index changes, and
is particularly useful for disentangling the superposition of components with 
different spectra that can overlap along the line of sight.

\begin{figure*}[]
\centering
\includegraphics[]{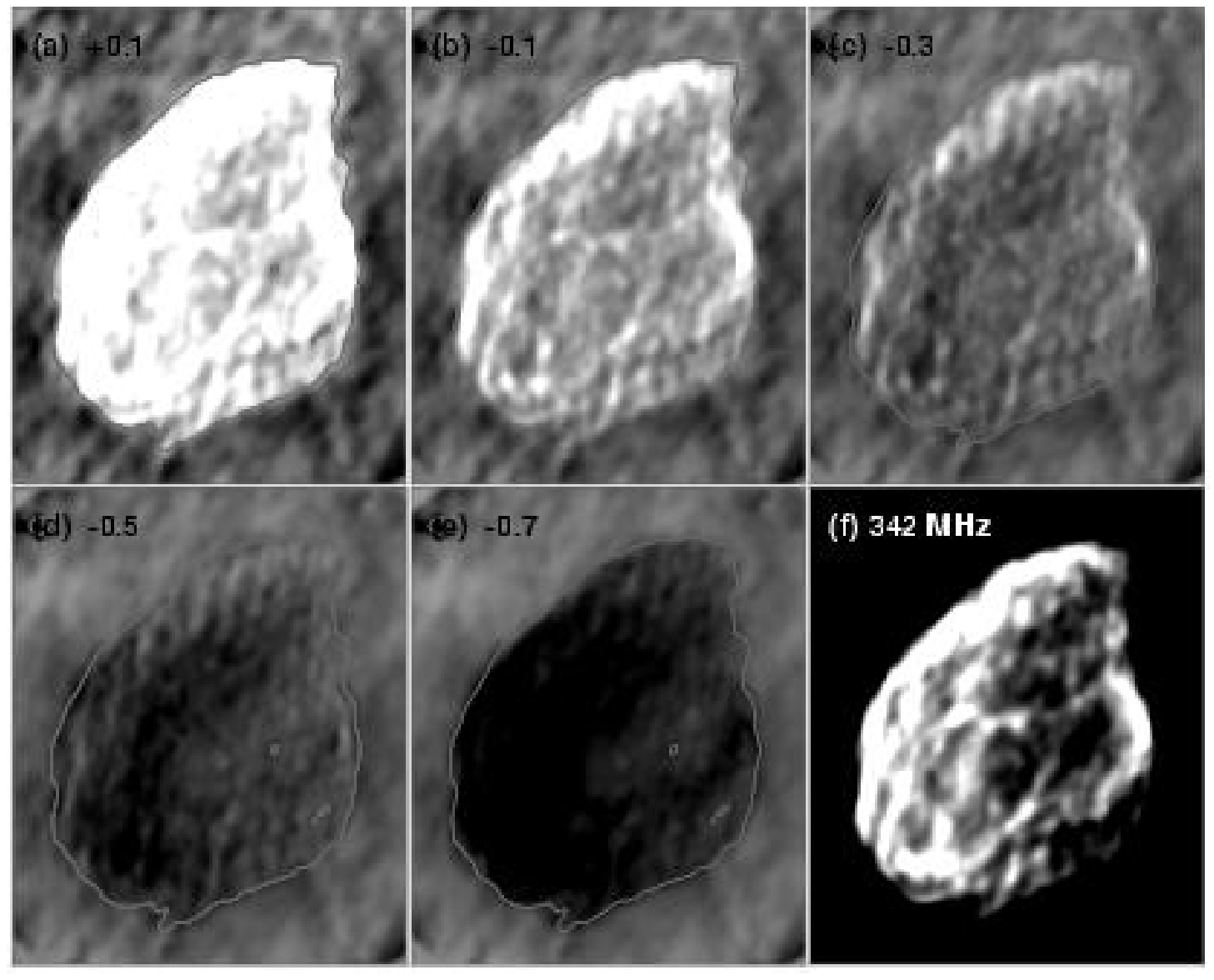}
\caption{
Gallery of tomographic images for the SNR \object{W44}
constructed from data at 324 and 1442 MHz, which were matched in uv
coverage.  The data at both frequencies have a common 50$\arcsec$ resolution.
The spectral index $\alpha_{\mathrm{t}}$ is indicated at the top left corner
of the panels.  Bright areas have radio continuum spectral indices steeper 
than the assumed $\alpha_{\mathrm{t}}$ and dark areas imply flatter spectrum.
The grayscale is the same in all panels.  A 324
MHz intensity contour at 260 mJy~beam$^{-1}$ is plotted for reference.
The image of \object{W44} at 324 MHz plotted for comparison is
displayed in the bottom right frame.
\label{tomo324-1442}}
\end{figure*}

\begin{figure*}[h]
\centering
\includegraphics[]{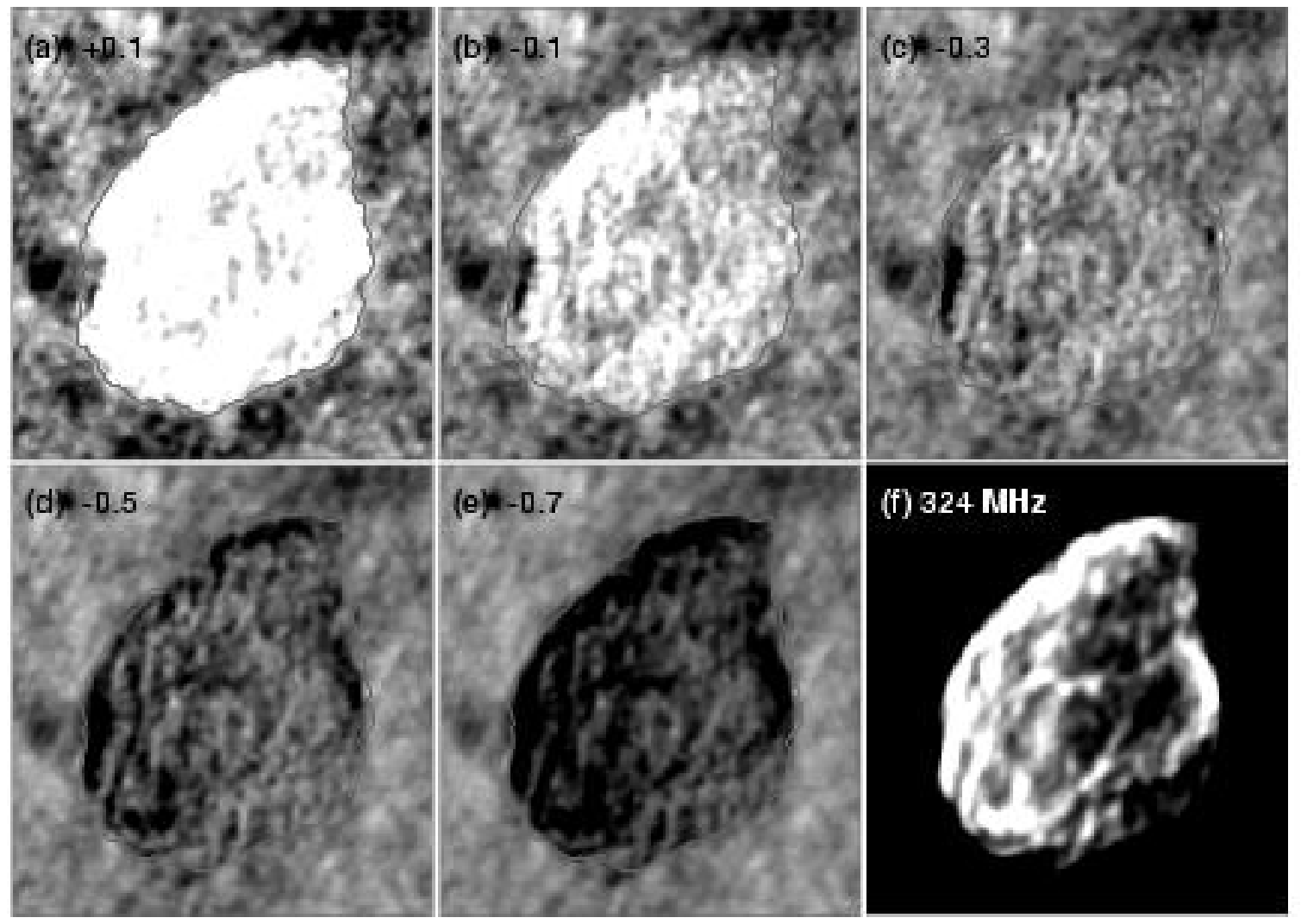}
\caption{The same as Fig.~\ref{tomo324-1442} but between 74 and 324
MHz. Bright areas
have radio spectral indices steeper than the assumed
$\alpha_{\mathrm{t}}$ and dark areas imply flatter spectral indices.
The greyscale is the same for each spectral index panel. The 324
MHz contour at 140 mJy~beam$^{-1}$ is also plotted in each panel for
reference.  A greyscale 324 MHz image with a range of 26 to 70
mJy~beam$^{-1}$ is shown in the lower right panel for comparison.
\label{tomo74-324}}
\end{figure*}

In Fig.~\ref{tomo324-1442} we present a tomographic
gallery of 324 vs. 1442 MHz data, while the comparison between 74 and
324 MHz data is shown in Fig.~\ref{tomo74-324}.  In both cases the
tomographic images were generated for five test spectral index
values:
$\alpha_{\mathrm{t}}$=+0.1,
$\alpha_{\mathrm{t}}$=$-$0.1,
$\alpha_{\mathrm{t}}$=$-$0.3,
$\alpha_{\mathrm{t}}$=$-$0.5, and $\alpha_{\mathrm{t}}$=$-$0.7.
In these maps the positive spatial components (seen as light regions) 
are steeper in
spectral index than the assumed test value, while the negative (dark)
residuals correspond to features having a spectrum flatter than the
test value.  In both figures, the bottom right frame
displays the 324 MHz continuum image for comparison.

In the 74/324 MHz comparison (Fig.~\ref{tomo74-324}), the feature with 
positive spectrum noted earlier
near $\sim\,18^{\mathrm{h}} \, 57^{\mathrm{m}}$, $01^{\circ}\, 17^{\prime}$
is strikingly revealed as a short, dark filament in the eastern border of 
all the tomographic images. It is especially evident in panels from
$\alpha_{\mathrm{t}}$=$-$0.1 to $\alpha_{\mathrm{t}}$=$-$0.3 as a structure 
with flatter spectrum than these values.  Also,
the tomographies traced between 74 and 324 MHz emphasize
the westernmost arc, that stands out as a flatter spectral index feature in the
images traced at $\alpha_{\mathrm{t}}$=$-$0.5 and $\alpha_{\mathrm{t}}$=$-$0.7 
and disappears against the background at $\alpha_{\mathrm{t}}$=$-$0.3, while 
for the tomographies traced between 324 and 1442 MHz (Fig.~\ref{tomo324-1442}) 
the best fit is attained for $\alpha_{\mathrm{t}}$ between $-$0.5 and $-$0.7, 
thus confirming the negative (concave down) curvature of the spectrum 
recognized earlier.

In summary, all of the important radio continuum spectral
index characteristics discussed in the previous section are confirmed through 
this alternative analysis.

An important conclusion from this new radio continuum spectral study is that 
neither spectral index maps nor tomographies reveal any trace of coupling 
between the pulsar and the surrounding plasma beyond the  
PWN; that is, there is no spectral signature, such as 
a smooth radial steepening, that would suggest a progressive aging of 
particles from their
injection near the pulsar to their transport further out into the SNR shell. 
Moreover, the spectral index
obtained for the shell ($\alpha$$\sim\,$$-$0.5), compatible with the diffusive 
shock acceleration mechanism in the test particle limit, confirms that the
relativistic electrons responsible for the shell radio emission are accelerated
right at the
shock front, with no influence from the central neutron star. 
Also, no spectral signature 
is observed that can be related to the presence of the 95\% confidence ellipse 
corresponding to the EGRET $\gamma$-ray source that covers most of the 
southeast corner of \object{W44}.

\section{Multi-wavelength Comparisons}
\subsection{Radio, infrared and molecular emission}

\begin{figure}[h]
\centering
\includegraphics[]{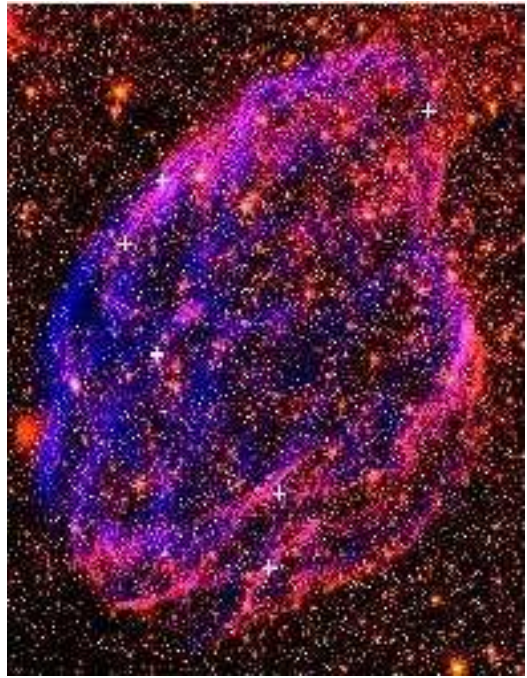}
\caption{
A color composite image showing the spatial correlation between the
mid-infrared emission as observed by the \it Spitzer Space Telescope
\rm at 4.5 $\mu$m (in orange) from \citet{rea06}, and the new low
frequency radio image at 324 MHz (in blue).  Features where both
spectral bands overlap are magenta in color.  The positions where 
OH (1720 MHz) maser emission was detected are indicated with 
white $+$ symbols \citep{hof05}.
\label{ir-radio}}
\end{figure}

Recently, \citet{rea06} used {\em Spitzer Space Telescope} GLIMPSE
data to resolve the mid-infrared structure of \object{W44} between 3.6
and 8 $\mu$m \citep{ben03}.  To study the relationship between the infrared 
and radio
features in the SNR \object{W44} we have combined the data at 
4.5 $\mu$m with the new radio image at 324 MHz in a false color image 
(Fig.~\ref{ir-radio}). 
In this figure, the mid-infrared emission is shown 
in orange
and the radio emission in blue; magenta thus corresponds to
regions where both spectral bands overlap.  The distribution of
compact OH (1720 MHz) maser spots \citep{cla97,hof05} are shown as plus signs.

The mid-IR (MIR) image reveals extensive filamentary emission and some
isolated bright clumps.  This emission is particularly bright in the
north and southwestern parts of \object{W44}, where bright near
infrared H$_{2}$ line (2.12 $\mu$m) emission is also present
\citep{rea05}.  
Figure~\ref{ir-radio} reveals the impressive agreement
between the low frequency radio emission and the infrared 4.5 $\mu$m
emission to the north and southwest, where the brightest radio
filaments are
almost exactly coincident with the infrared radiation.  In contrast,
the central and eastern regions do not show much obvious correspondence
with the exception of a few bright knots.  The westernmost bright
radio arc near
$18^{\mathrm{h}}\,55^{\mathrm{m}}\,20^{\mathrm{s}}$
$01^{\circ}\,22^{\prime}$, which is also
bright in H$_{\alpha}$, [SII] and [NII] lines \citep{gia97, mav03}, 
has an infrared counterpart that mimics the appearance of the radio arc.

\begin{figure*}[h]
\centering
\includegraphics[]{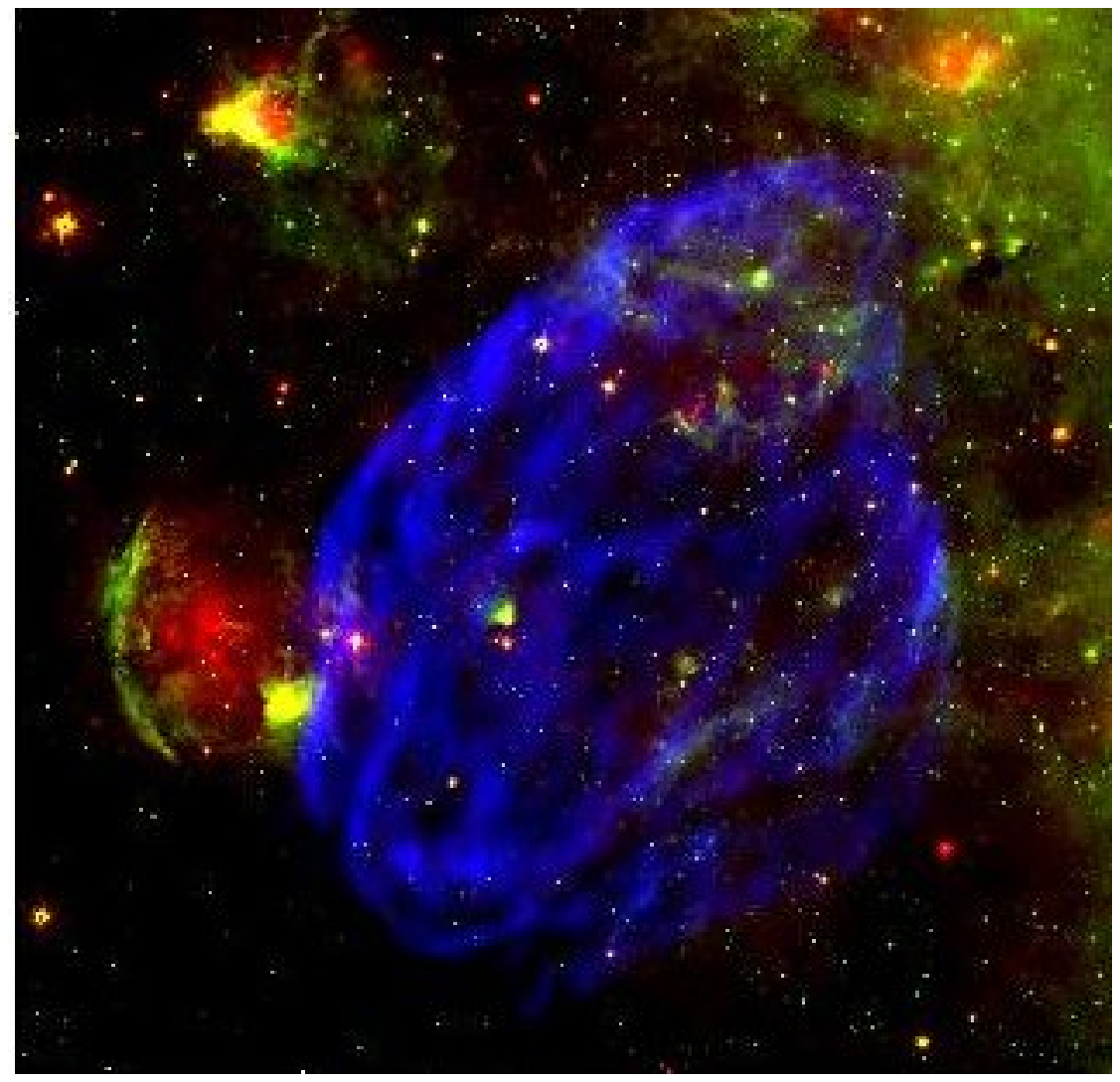}
\caption{A high resolution comparison of the radio continuum emission
at 324 MHz of the \object{W44} and the infrared emission in the SNR
region. The green and red images corresponds to the \it Spitzer \rm
8 and
24 $\mu$m data, while in blue the 324 MHz image of the remnant with
13$\arcsec$ resolution is shown.
\label{infrared}}
\end{figure*}

The mid-IR study is also useful for understanding \object{W44} in the context 
of the surrounding interstellar medium. Figure~\ref{infrared} displays 
a three color image of W44 comparing the radio morphology at 
324 MHz (blue) with {\em Spitzer} GLIMPSE and MIPSGAL images at 8 $\mu$m 
(green) and 24 $\mu$m (red),  respectively (Carey et al., in preparation). These
infrared bands are excellent tracers of the warm
dust associated with star formation. 
The most conspicuous feature in Fig.~\ref{infrared} is the $\sim\,$10 arcmin 
region
just outside the southeast limb of \object{W44} (red nebula), identified by
\citet{pal03} as the HII region \object{G034.7$-$00.6}. In this case the 
combination of the 8 and 24 $\mu$m images reveals with impressive detail the 
hot dust grains (in red) in the Str\"{o}mgren sphere, limited to the east by 
an annular photo dissociation region (PDR) dominated by
polycyclic aromatic hydrocarbon (PAH) emitting near 8 $\mu$m (green). The 
bright yellow spot abuts the only portion of the eastern limb of \object{W44} 
with a concave shape 
(near $\sim\,$$18^{\rm h}57^{\rm m}$, $01^{\circ}\, 17\arcmin$) 
coincident with the brightest radio filament and the region determined to have
a positive (inverted) radio spectrum (as shown in \S4.2). The IR point source 
is the IRAS source
\object{18544+0112}, proposed by \citet{mol96} to be a young stellar object 
(YSO) in a primitive evolutionary stage.

As summarized in the introduction (\S1),
\object{W44} is one of the few SNRs known to be physically
interacting with its parent molecular cloud complex, with high-angular 
resolution molecular studies in
different near-IR (NIR) and millimetric lines
\citep[see for example][]{set04,rea05}
clearly indicating its evolution within a giant molecular cloud.
In particular, one molecular cloud component has been detected adjacent and 
closely parallel to its bright eastern radio boundary \citep[][see for 
instance Fig. 3e in their paper]{rea05},
while another one is located near the very bright western arc. \citet{set04} 
provide a useful schematic illustration of the various CO features discovered 
towards \object{W44} (Fig. 3 in their paper). In particular, and near the 
radial velocity of $\sim\,$47 km~s$^{-1}$ approximately corresponding to the 
2.9 kpc distance attributed to \object{W44}, they identify a CO
``edge'' feature exactly at the position of the YSO and the indentation of the
eastern rim (E2 in their nomenclature). Based on the abrupt changes observed 
in the
physical parameters, \citet{set04} interpret the SNR as interacting with this 
particular molecular
cloud component.

In this scenario involving molecular gas, an SNR, an HII region, and a YSO, it 
is worthwile to ask whether this may be a case of star formation triggered by 
the action of a SN shock on a coincident molecular cloud or, alternatively, 
whether it might illustrate a case of
sequential star formation at the periphery of the HII region. This last case
has been, for example,
observed in the \object{Sh~217} and \object{Sh~219} HII regions \citep{deh03}.
Although \citet{mol96} proposed a kinematic distance of 3.7 kpc for the YSO, 
this distance might be an overestimate and the YSO is instead
co-distant with \object{W44}, at about 2.9 kpc. The question then is 
whether the age of
\object{W44}, of a few 10$^{4}$ yrs, is sufficient to have been responsible 
for the formation
of new, massive stars. The timescale of star formation is currently under ample
debate, but it is generally agreed that a few 10$^{6}$ yrs are necessary for the
formation of protostars from condensed molecular matter 
\citep[e.g.][and references
therein]{tas04}. We are therefore led to the conclusion that if the YSO is 
related
to \object{W44}, then the cloud collapse must have been initiated by the 
action of the 
precursor star's wind and not directly by the SNR. 
Therefore we feel it is more plausible that the YSO is associated with the PDR 
around the HII region
\object{G034.7$-$00.6}. Detailed NIR, MIR and millimeter studies are planned to
further investigate the star formation in this complex region, with the goal 
of uncovering the underlying genetic links
between the SN shock, the HII region, and the protostar.

From Fig.~\ref{infrared} regions of more
nebulous infrared emission toward the north-northwestern sides of \object{W44} 
are also apparent, 
and the yellow spot about $\sim\,$1 arcmin in size seen to
the east of its geometrical center corresponds to the HII region
\object{G034.7$-$00.5} 
\citep[centered at 18$^{\rm h}$ 56$^{\rm m}$ 26.5$^{\rm s}$,
01$^{\circ}$ 20$\arcmin$ 38$\arcsec$,][]{pal03}. Observations of radio 
recombination lines (RRL) would
clearly be helpful to further elucidate the presence of thermally emitting gas 
in this area. 

\subsection{Radio and X-ray emission}

\begin{figure*}[]
\centering
\includegraphics[]{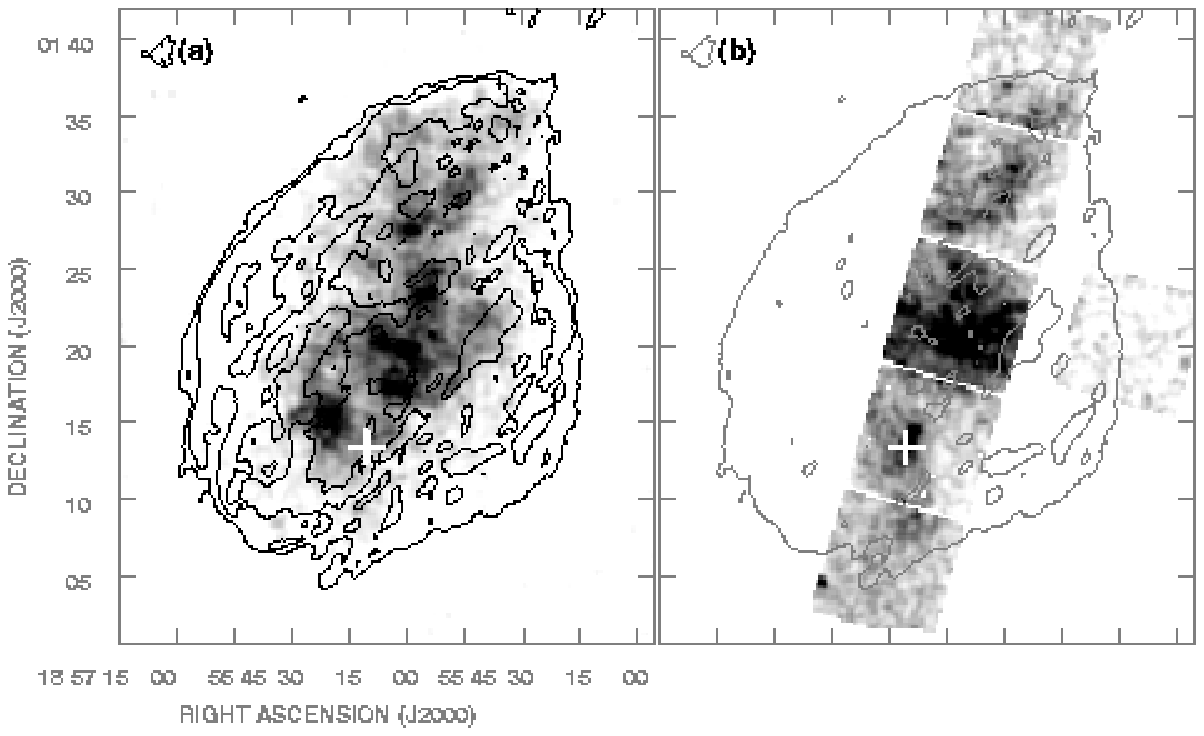}
\caption{
An X-ray/radio comparison of the SNR \object{W44}. \bf (a) \rm The
grayscale corresponds to the thermal X-ray emission observed with \it
ROSAT \rm in the 0.2-2.4 KeV range, while the overlaid contours trace
324 MHz radio emission at 20 and 50 mJy~beam$^{-1}$.  \bf (b) \rm 
\it Chandra \rm 0.7-2.6 keV image of \object{W44} 
\citep[as taken from][]{she04} with the 20 mJy~beam$^{-1}$ (4$\sigma$) radio 
CONTOUR at 324 MHz superimposed for comparison.
The white plus sign shows the position of the pulsar
\object{PSR~B1853+01}.
\label{xray-radio}}
\end{figure*}

The X-ray emission from \object{W44} has been investigated using the
\it Einstein\rm, \it EXOSAT\rm, \it ROSAT\rm, \it ASCA\rm, and \it
Chandra \rm telescopes \citep{smi85,jon93,rho94,har96,she04}.  In
Fig.~\ref{xray-radio}a we present a comparison between the 324 MHz
radio image and the X-ray emission as observed with \it ROSAT \rm in
the energy range 0.2 to 2.4 keV (taken from \it ROSAT \rm
archives). As reported by \citet{rho98}, the brightest X-ray
features are preferentially located in the center of the remnant, in a
region of relatively low radio surface brightness.  As expected, no
correspondence is observed between the mostly thermal X-ray emission 
\citep{she04} and the filamentary synchrotron radio emission.  A
number of physical processess such as cloud evaporation, electron
thermal conduction, and entropy mixing have been proposed to explain
the observed X-ray properties \citep[see][for a review of the
models]{she04}.  \citet{rea05} conclude that probably both,
evaporating clumps and thermal conduction behind radiative shocks, are
operating to produce the observed X-ray radiation.

Fig.~\ref{xray-radio}b shows the \it Chandra \rm image of \object{W44}
produced by \citet{she04} from the combination of 6 different pointings 
with the outer
contour (traced at 4$\sigma$ level) of the 324 MHz radio emission
plotted for comparison.  \citet{she04} suggest that diffuse X-ray
emission is observed extending slightly beyond the northwestern
border of the radio synchrotron emission.  The new high-sensitivity
radio images presented here, however, do not support this conclusion.

\section{Energetics of SNR~\object{W44}}
As \object{W44} harbours a pulsar with a PWN and is spatially coincident with 
the GeV gamma-ray source \object{3EG~1856+0114}, it is useful to calculate its 
total energy content and compare it with other Galactic SNRs with central
compact objects.
The total synchrotron energy is
stored in the magnetic field as well as in the kinetic energy of the
relativistic particles.  
Under the assumption of equipartition between particles and magnetic energy, 
the minimum energy  content and the corresponding magnetic field can be 
obtained from the following relations \citep{mof75} 

\begin{equation}
U_{min}=0.50\, (a\,A\,L)^{4/7}\,V^{-3/7}
\label{energy}
\end{equation}

\begin{equation}
B_{min}=2.3\,(a\,A\,L/V)^{2/7}
\end{equation}

where $V$ is the volume of the source, $A$ is a factor
which depends on the assumed lower and upper cutoff frequencies and
the spectral index, and $L$ is the total luminosity of the
source. We note that electrons
may not be the only energetic particles within the source; an
appreciable amount of energy may also be stored in energetic baryons.
The total particle energy can be described as $U_{p}$=$a\,U_{e}$,
where $a$, the energy ratio between electrons and baryons
is usually assumed to be $\sim$50 \citep{mof75}. 

Using the integrated spectral index of $-$0.37 and assuming isotropic
radiation, we derive a total luminosity for \object{W44} of 7.8 $\times$ 
10$^{34}$ erg~s$^{-1}$. The volume of the source is estimated to be about 
2 $\times$ 10$^{59}$ cm$^{3}$ by approximating \object{W44} as a 
sphere of radius 11.6 pc. 
For typical lower and upper cutoff frequencies
of 10$^{7}$ and 10$^{11}$ MHz, the total minimum energy content is 
5.8 $\times$ 10$^{49}$ erg. Following the same procedure and with similar 
assumptions we also re-estimated the total energy content of the PWN based on 
the 324 MHz data.
We derive its minimum energy as $U_{min}(PWN)=1.2\,\times\,10^{47}$ 
erg, in good agreement with \citet{fra96}'s results 
($U_{min} \simeq 1 \,\times\,10^{47}$ erg) estimated from higher radio 
frequencies observations. 

The minimum energy derived for \object{W44} is comparable to the value 
obtained for SNR \object{CTB80} 
\citep[$U_{min}=5\,\times\,10^{49}$ erg,][]{cas03}, 
another SNR containing a pulsar, and two orders of magnitude higher 
than the energy content of \object{W28} \citep[$U_{min}=2\,\times\,10^{47}$ 
erg,][]{dub00}.

The corresponding minimum magnetic field in the SNR is 
$B_{min}$=18 $\mu$G, consistent with the compression of a typical ISM 
magnetic field  
\citep[about 5 $\mu$G,][]{wie05} by a factor of three to four.

\section{Conclusions}
We have presented new high-sensitivity and
high-resolution low frequency VLA radio images of the SNR \object{W44}.
These images provide an unprecedented view of \object{W44} with
an angular resolution of $\sim$37$^{\prime\prime}$ and 
$\sim$13$^{\prime\prime}$
at 74 and 324 MHz, respectively, an improvement of over two
orders of magnitude over the best, previous low-frequency image 
of this SNR.

Based on our 74 MHz image sensitive to structures with angular scales 
from a few arcseconds to $\sim$ half a degree, we measure an integrated flux
density for \object{W44} of S$_{74\,\mathrm{MHz}}\simeq\,$634 Jy. 
Similarly, from our 324
MHz image sensitive to spatial structures between $\sim$13 arcsec and $\sim$1
degree, we derive an integrated flux density 
S$_{324\,\mathrm{MHz}}\simeq\,$411 Jy.
We have used these measurements, together with carefully considered flux 
densities
present in the literature, to calculate an accurate global
radio spectrum of the SNR.  The best fit to the data yields an integrated
spectral index $\alpha$=$-$0.37$\pm$0.02, constant over four decades in 
frequency
with no significant turnover down to 74 MHz. The straight power law spectrum is
in contrast to a number of other SNRs that exhibit spectral turnovers at low 
frequencies,
and is consistent with previous low frequency measurements of this object. 

The new 74 and 324 MHz data, together with existing 1442 MHz data,
were combined to perform the first careful, spatially resolved
study of the synchrotron spectrum in \object{W44}.
High-dynamic range observations at three well spaced frequencies are
ideal for accurately constraining the variation of the spectrum
within the 
source. Our results were obtained first from standard analysis of spectral 
index maps, and thereafter confirmed using spectral tomography techniques.  
The main findings of our radio continuum spectral index study are:

(i) The eastern limb, which is the brightest portion of \object{W44} SNR, 
as well as the narrow filaments observed in total power across most of the 
surface of the source have a straight power law spectrum between 74 and 1442 
MHz with $\alpha$$\sim\,$$-$0.4, $-$0.5. This result is consistent with the 
predictions of a simple, diffusive shock acceleration model applied to the 
filamentary shell of \object{W44}.

(ii) From the 74/324 MHz comparison we have discovered a narrow feature
with a positive spectrum (0.0 $<$ $\alpha$ $<$ +0.4 ) located towards the 
southeast 
border of \object{W44}, coincident with an indentation in the limb of the SNR. 
At this location existing studies have shown that the SNR shock is interacting
with an external 
molecular cloud. Also, flanking this portion of \object{W44}, mid-IR 
observations have revealed the presence of an almost circular HII region 
surrounded by a PDR region. 
Exactly at the interface between this HII region and W44, as seen in the plane 
of the sky, a young stellar object is present. A likely explanation of the
inversion in the low frequency radio continuum spectrum in this region 
is free-free absorption from
ionized gas in the post shock region at the SNR/molecular cloud
interface. If this is the case, it marks the second case of
spatially resolved low frequency
radio observations tracing an SNR/molecular cloud interaction, following 
\citet{bro05}'s 
detection of thermal absorption at the periphery of 
\object{3C~391}. An alternative explanation is that the thermal absorption is 
occuring at the borders of the cavity carved by the collindant HII region,
where the column
density of thermal electrons might be highest. The 
spectral inversion is probably produced by a combination of both these effects. 
However this localized absorption is negligible compared to the total 
integrated flux
and thus has no measurable influence on the integrated continuum spectrum.
Further IR and 
millimeter observations are planned to investigate the 3-D distribution of the 
gas emitting in the different spectral regimes, and to elucidate the 
precise physical processes 
acting at this site. This should shed light on the complex chain of 
interconnected events
that can tie together the evolution of \object{W44}, its parent molecular 
cloud, precursor star,
and neighbouring HII region in a scenario consistent with the formation of a 
new generation
of stars exactly at the interface SNR/molecular cloud.  

(iii) The comparison of the spectrum between 74 and 324 MHz and between 324 and 
1442 MHz revealed a spectral flattening ocurring on the westernmost arc of 
\object{W44} at the lowest frequency pair (concave down spectrum). 
Our hypothesis
is that this spectral flattening is related to the SNR shock colliding with a 
molecular
cloud in this region, consistent with the bright optical filaments and IR 
emission observed there. 
In this scenario the spectral flattening could be a consequence of strong 
postshock
densities and enhanced magnetic fields. If the electron energy spectrum 
hardens at lower frequencies, regions in which the magnetic field is higher 
than the surroundings
will appear both brighter and with a flatter spectrum. A firm estimate of the 
distribution of
magnetic field strength throughout the SNR would be important to test this 
hypothesis. 

(iv) The present study  revealed that there is no evidence in the radio 
continuum spectrum of any coupling between the associated pulsar 
\object{PSR~B1853+01} and the surrounding 
SNR shell (that could, for example, have been observable as a gradual 
steepening from the pulsar to the shell). 
Furthermore, there is no spectral signature
indicating any link to the EGRET $\gamma$-ray source whose 95\% 
confidence ellipse
covers most of the southeast corner of \object{W44}. 

We have also used the new 324 MHz image to perform the first
detailed low frequency radio/IR comparison for \object{W44}.  Bright
filamentary IR emission detected by \it Spitzer \rm  at
4.5 $\mu$m shows an impressive correspondence with the radio emission 
towards the north and western portions of the remnant. To the central and 
eastern regions 
of \object{W44}, however, almost no
IR counterpart is observed for the radio emission.

From the comparison of the new radio image at 324 MHz with the X-ray
emission as seen by \it ROSAT \rm between 0.2 and 2.4 keV and with a
high-resolution \it Chandra \rm image, we can confirm that the
synchrotron radio emission appears to surround the thermal X-ray
plasma. As determined from previous studies, the X-ray emission is probably
related to a combination of the evaporation of clouds and thermal 
conduction
occuring behind radiative shocks.  In addition
heavy absorption likely produced by foreground molecular clouds
may explain the absence of X-ray radiation towards the
southeast.

We have also investigated the energetics in \object{W44} based on the new 
observations and accurate, re-derived integrated spectrum. We find that the 
total minimum energy content is 
5.8$\times$10$^{49}$ erg, a value in agreement with the estimates for other 
SNRs 
containing central pulsars. The minimum, equipartition magnetic field estimate 
is 18 $\mu$G. 

Finally, we note that this study adds to a growing list of examples confirming
the importance of high resolution, low frequency radio observations 
in providing an
important new tool to demonstrate and understand the physical interaction 
between SNR shocks and the surrounding dense material. This also points to the
great potential of an emerging generation of much more sensitive low frequency
instruments.

Future work will involve a detailed comparison of the radio
spectral index distribution and the molecular emission in order to refine the 
analysis of the interaction between \object{W44} and its parent molecular 
cloud. New 
IR and optical studies are also planned in order to improve our understanding
of the underlying genetic links and causal relationships between the 
evolution of the SNR and the adjacent YSO in this inner Galactic complex.

\begin{acknowledgements}
We thank W. Reach for supplying us with the IR data in electronic form.
We also acknowledge the referee for his constructive comments.
This research has been funded by Argentina grants ANPCYT-PICT 04-14018,
UBACYT A055/04, ANPCYT-PICT 03-11235 and PIP-CONICET 6433. Basic
research in Radio Astronomy at the Naval Research Laboratory is
supported by 6.1 base funding. Data processing was carried out using
the HOPE PC cluster at IAFE.
\end{acknowledgements}
\bibliographystyle{aa}  
\bibliography{bib-w44}
\IfFileExists{\jobname.bbl}{}
{\typeout{}
\typeout{****************************************************}
\typeout{****************************************************}
\typeout{** Please run "bibtex \jobname" to optain}
\typeout{** the bibliography and then re-run LaTeX}
\typeout{** twice to fix the references!}
\typeout{****************************************************}
\typeout{****************************************************}
\typeout{}
}

\end{document}